\documentclass{article}

\usepackage[nonatbib, preprint]{neurips_2022}

\usepackage[utf8]{inputenc} % allow utf-8 input
\usepackage[T1]{fontenc}    % use 8-bit T1 fonts
\usepackage{hyperref}       % hyperlinks
\usepackage{url}            % simple URL typesetting
\usepackage{booktabs}       % professional-quality tables
\usepackage{amsfonts}       % blackboard math symbols
\usepackage{nicefrac}       % compact symbols for 1/2, etc.
\usepackage{microtype}      % microtypography
\usepackage{xcolor}         % colors
\usepackage{lipsum}  

\usepackage{graphicx}
\usepackage[caption=false]{subfig}
\usepackage{biblatex}
\usepackage{xcolor}

\title{Domain Agnostic Pipeline for \\ Retina Vessel Segmentation}

\author{%
  Benjamin Hou \\
  Biomedical Image Analysis, \\ Imperial College London, UK \\
  \texttt{bh1511@imperial.ac.uk}
}

\begin{document}

\maketitle

\begin{abstract}
Automatic segmentation of retina vessels plays a pivotal role in clinical diagnosis of prevalent eye diseases, such as, Diabetic Retinopathy or Age-related Macular Degeneration. Due to the complex construction of blood vessels, with drastically varying thicknesses, accurate vessel segmentation can be quite a challenging task. In this work we show that it is possible to achieve near state-of-the-art performance, by crafting a careful thought pre-processing pipeline, without having to resort to complex networks and/or training routines. We also show that our model is able to maintain the same high segmentation performance across different datasets, very poor quality fundus images, as well as images of severe pathological cases. Code and models featured in this paper can be downloaded from \url{http://github.com/farrell236/retina_segmentation}. We also demonstrate the potential of our model at \url{http://lazarus.ddns.net:8502}.
\end{abstract}

\section{Introduction}

\textbf{Background}: Retinal illnesses, such as Diabetic Retinopathy (DR) or Age-related Macular Degeneration, are one of the most common causes of blindness. About 2.2 billion people worldwide~\cite{steinmetz2021causes} suffer from some form of eye disease. Early detection plays a pivotal role in clinical diagnosis, as retina degeneration is unidirectional and early treatment can decelerate further degradation.

Various eye related illnesses, such as Age-related Macular Degeneration (AMD) or Diabetic Retinopathy (DR), can be identified via images of retinal blood vessels~\cite{hong2022retinal}. Manually segmenting the blood vessels by hand is a time consuming and tedious task, thus it is desirable to expedite the process through automated means. The advantages of an automated solution brings consistency to the clinical workflow, higher throughput, and reduced human error due to factors such as being fatigued. 

Unlike CT or MRI modalities, where each pixel has a real world physical meaning, retina fundus images are taken from standard DSLR cameras. Different institutions may acquire images with drastically different imaging parameters, thus leading to datasets being highly domain sensitive.

\textbf{Related Works}: Vessel segmentation has been an age-old problem, spanning from methods that features traditional image processing techniques to modern deep learning methods. Traditional methods tends to be unsupervised, and relies more on image augmentation tools such as filters, wavelet transforms and thresholding (e.g. Kahn et al.~\cite{bahadarkhan2016morphological}). On the advent of deep learning, with the introduction of the UNet~\cite{DBLP:conf/miccai/RonnebergerFB15}, brought an influx of networks with varying architectures and complexity, such as; nested~\cite{DBLP:journals/pr/Zhao0WY21}, multiscale~\cite{DBLP:conf/isbi/FuXWL16}, fusion~\cite{DBLP:journals/inffus/YinCW22} and generative models~\cite{DBLP:conf/miccai/KamranHTZSB21}.

\textbf{Contribution}: In this work, we highlight the importance of formulating an appropriate machine learning pipeline for the task of vessel segmentation, to achieve near state-of-the-art performance whilst keeping the complexity of the network and training routine to a minimum. We demonstrate our method is on par with state-of-the-art models, as well as robust to domain shifts.

\section{Methods}

\begin{figure}[ht]
    \centering
    \includegraphics[height=1.6cm]{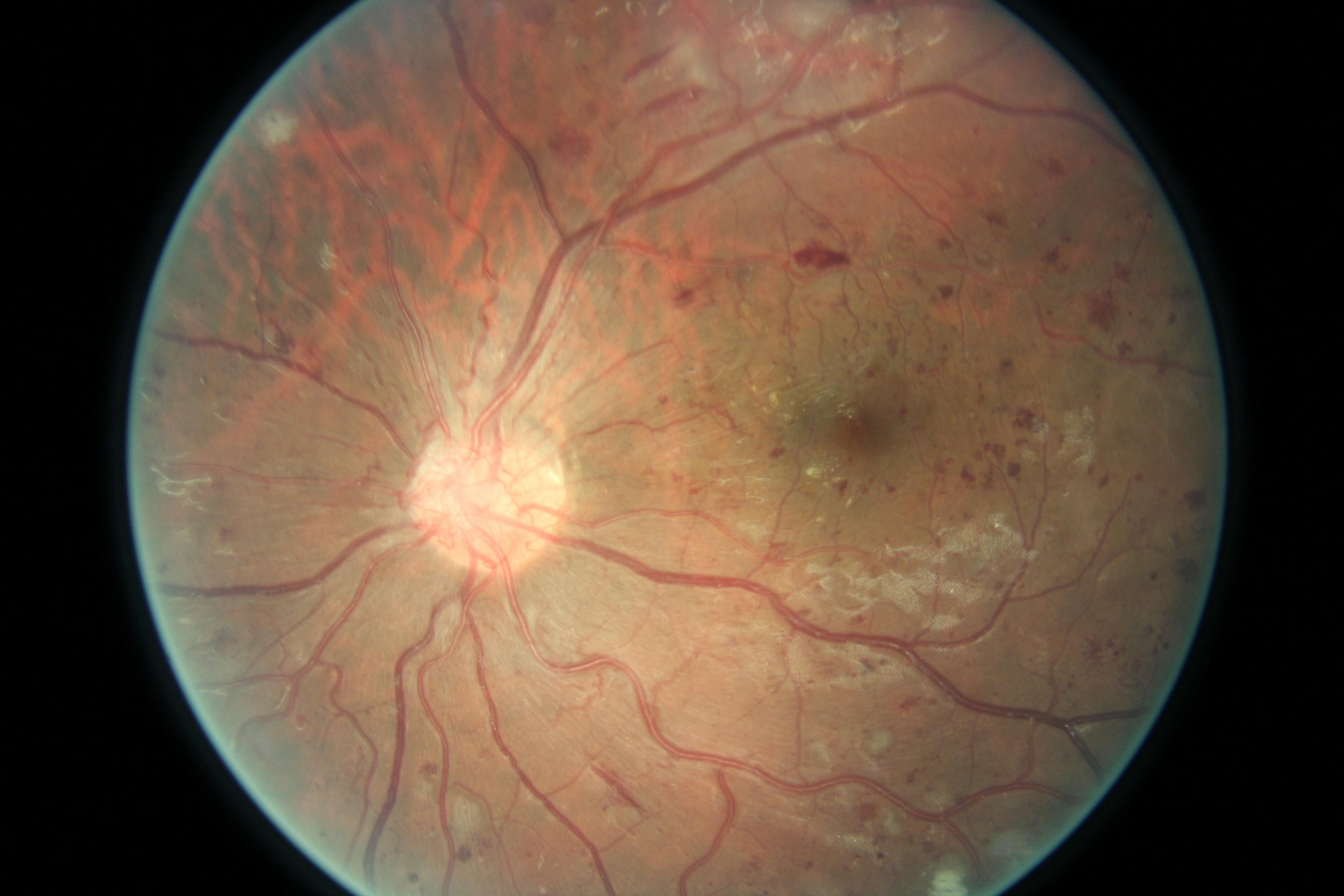} 
    \includegraphics[height=1.6cm]{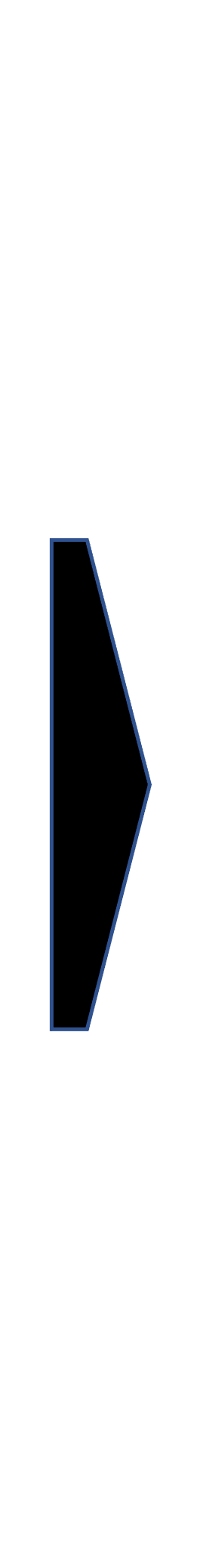} 
    \includegraphics[height=1.6cm]{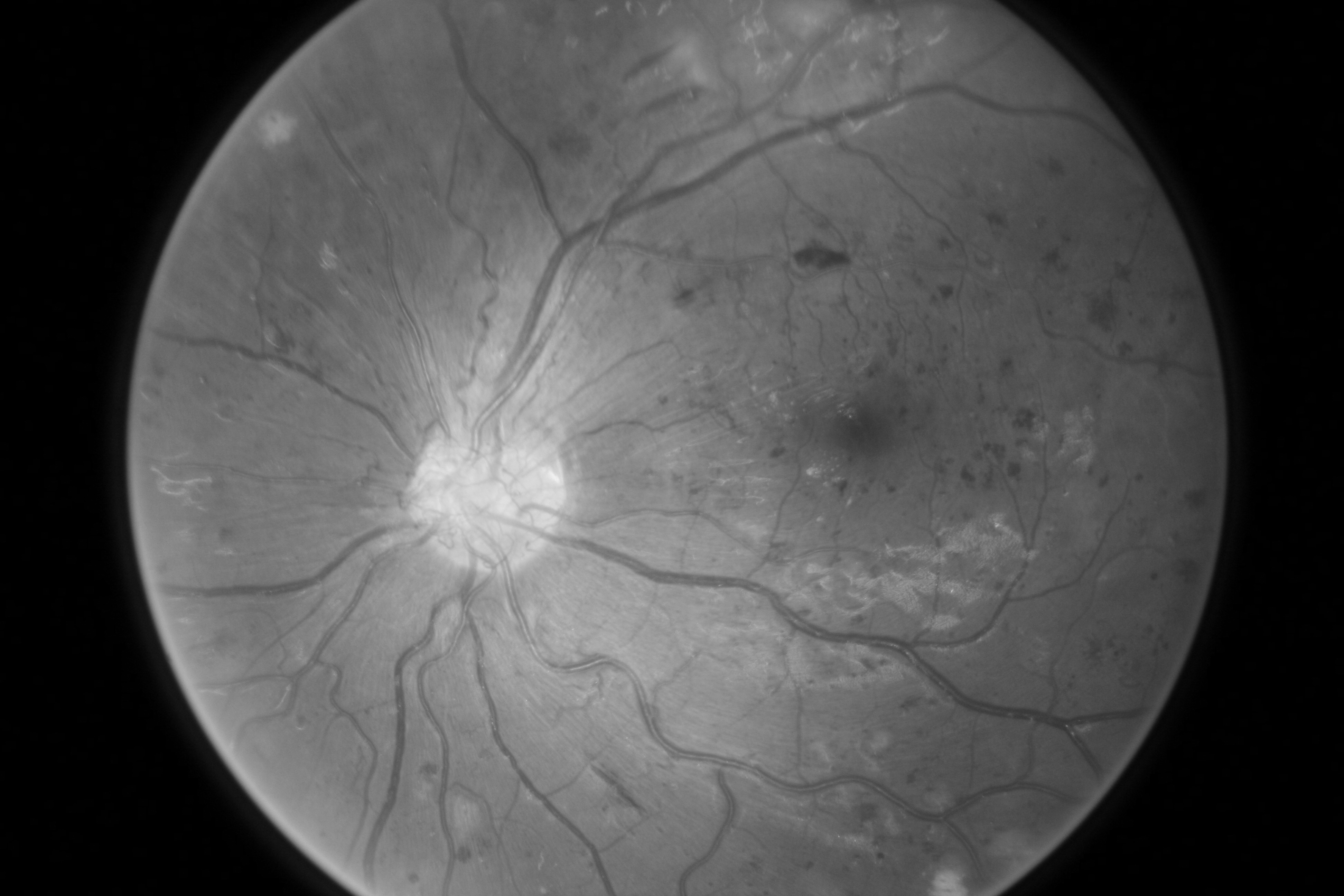} 
    \includegraphics[height=1.6cm]{arrow.png} 
    \includegraphics[height=1.6cm]{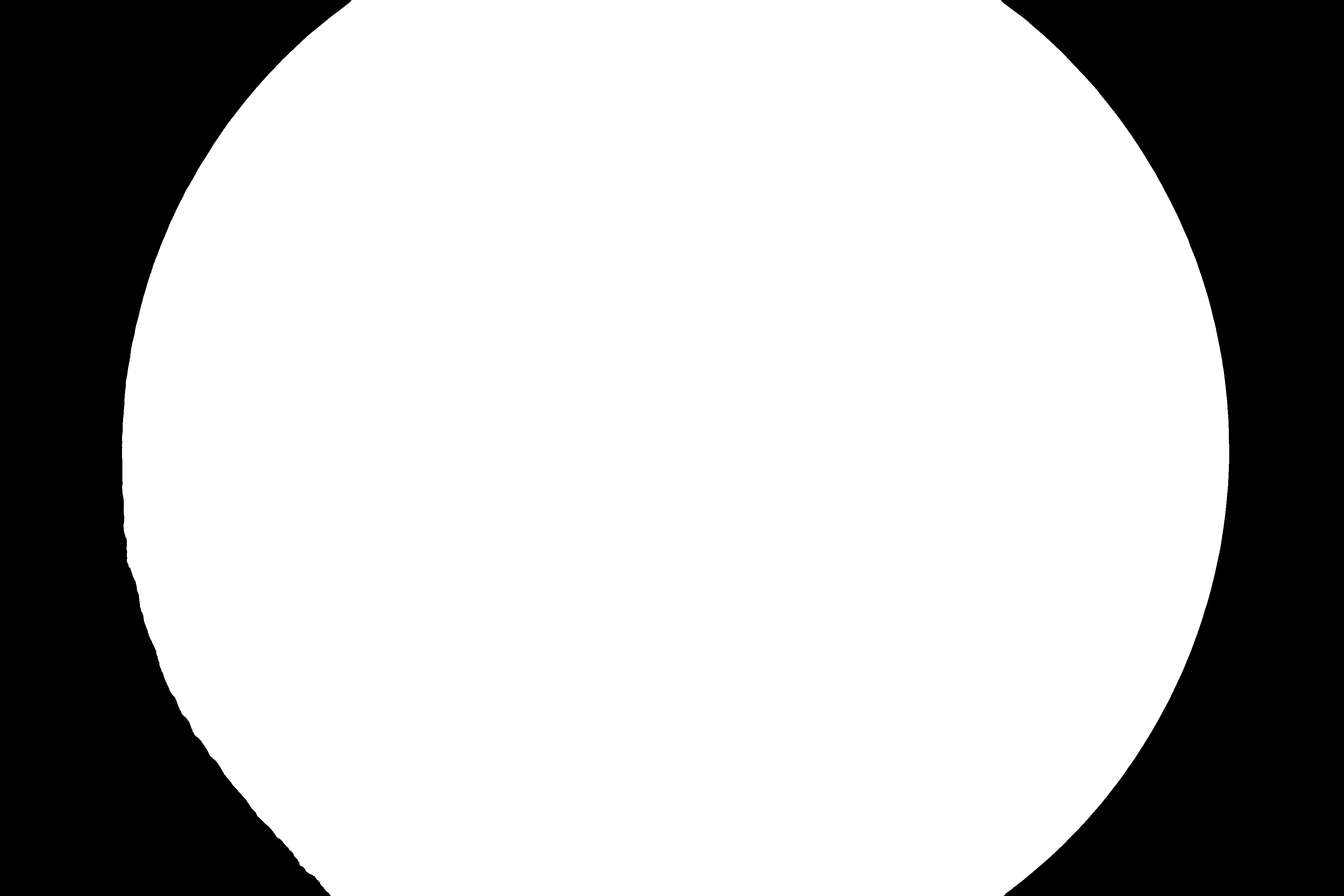}
    \includegraphics[height=1.6cm]{arrow.png} 
    \includegraphics[height=1.6cm]{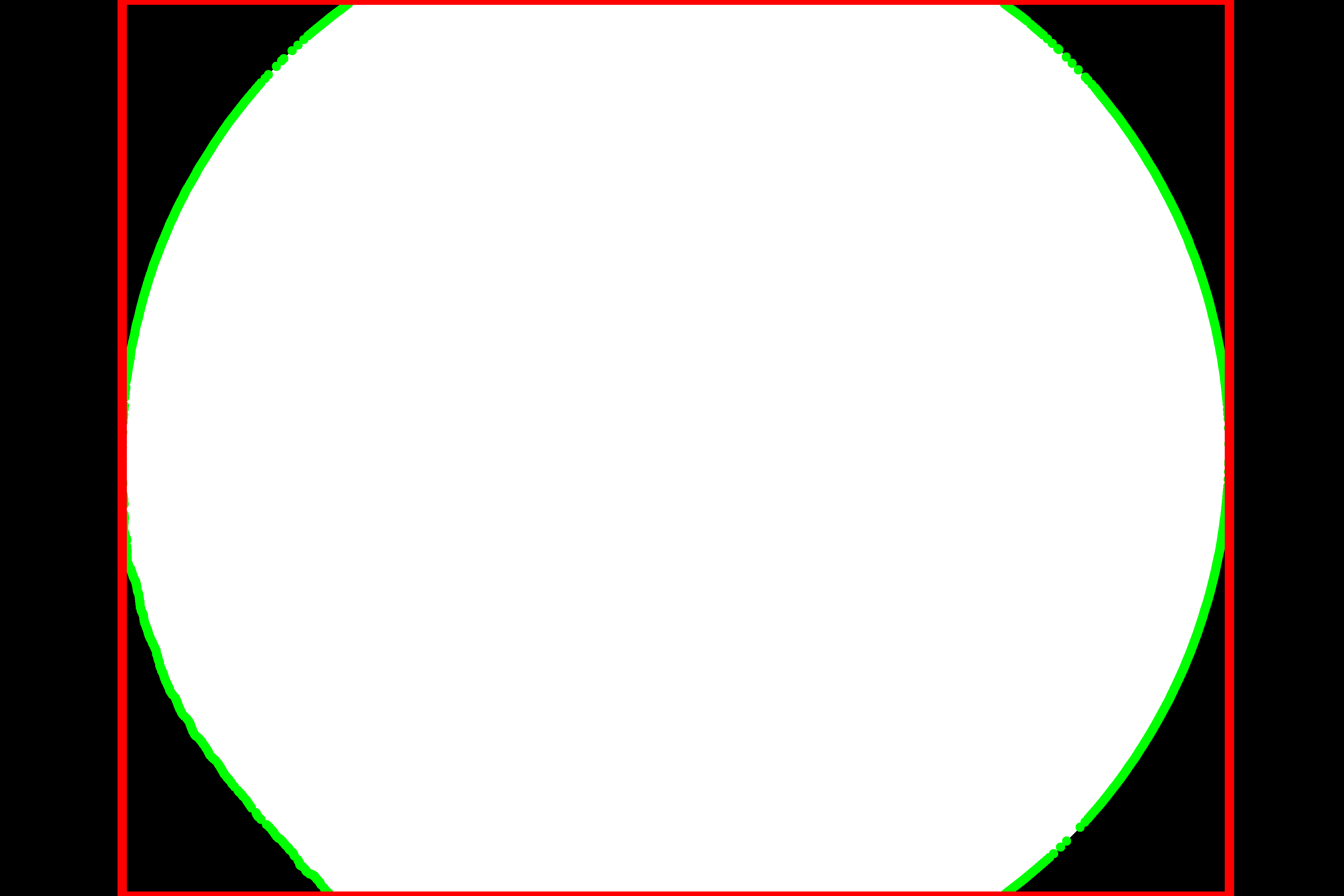}
    \includegraphics[height=1.6cm]{arrow.png} 
    \includegraphics[height=1.6cm]{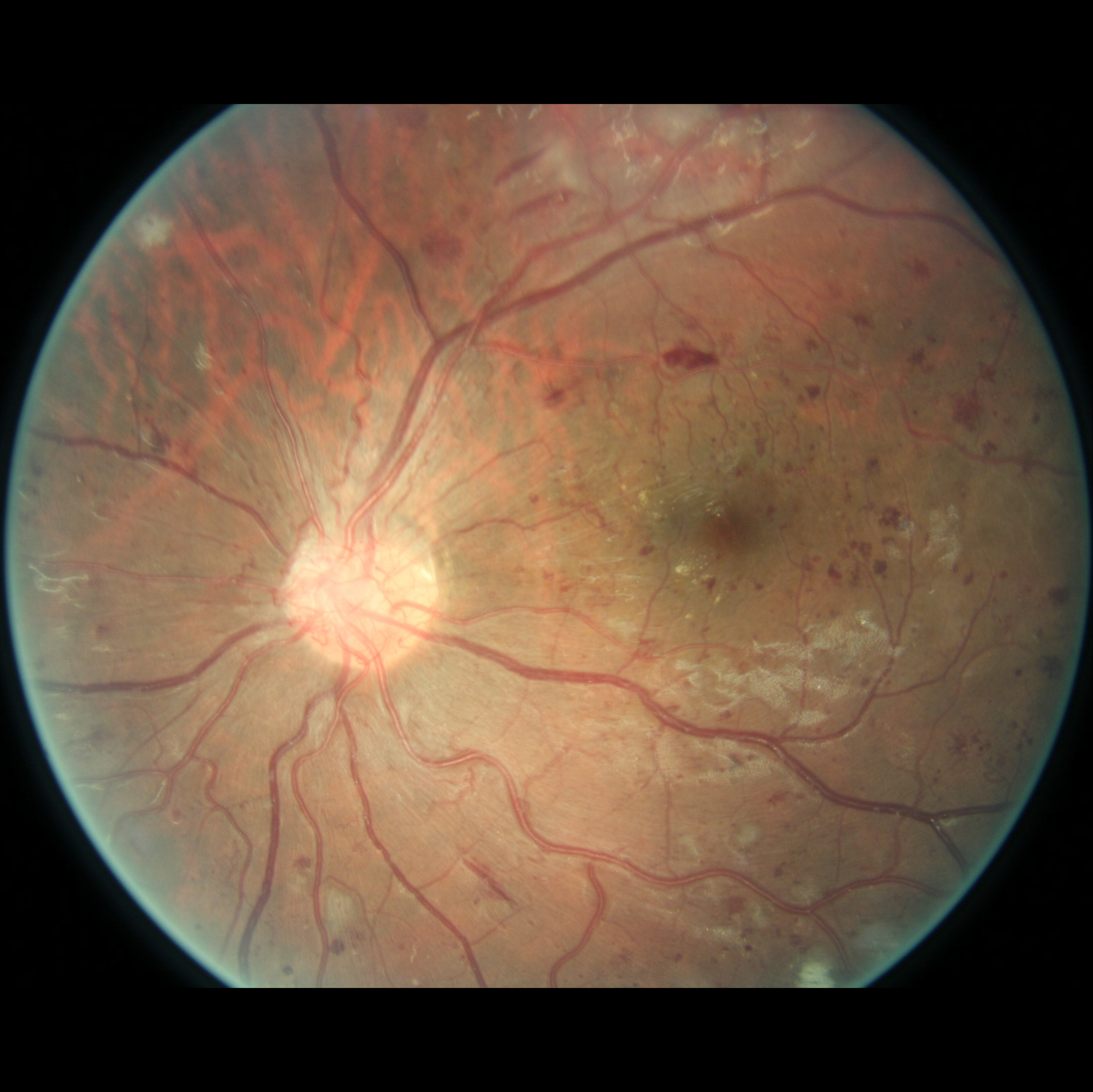}
    \caption{Retina Locator and Standardization} 
    \label{fig:pipeline}
\end{figure}

The pivotal focus of this work is to craft a careful and well-sought retina fundus image pre-processing and standardization pipeline to train generic deep learning models. At the start of the pipeline is the retina locator, which aims to locate the spherical vitreous body despite irregularities in shape and/or location in the raw captured photo. This is outlined by Figure~\ref{fig:pipeline}.

The captured retina photo is first turned into grayscale, where it is thresholded by 1/3 of the mean pixel intensity to get a preliminary mask. The mask is then cleaned by a median blur filter (ks=25), followed by 2 iterations of erode and 2 iterations of dilate to remove any speckle that may arise from the edges. Finally, the bounding box for the retina can then be inferred by the contour edge of the mask. Once the retina has been cropped and padded from the original photo to a square, it is then converted into grayscale, where Contrast-Limited Adaptive Histogram Equalization (CLAHE) is then applied. Adaptive Histogram Equalization is a family of methods that remediates over-amplification of the contrast. It operates on sub-regions of the image (known as tiles) to improve local contrast and enhance definitions of edges in each region, this very fitting for highlighting vessel-like structures.

\section{Experiments and Results}

Experiments were conducted on three publicly available vessel segmentation datasets; DRIVE~\cite{DBLP:journals/tmi/StaalANVG04}, STARE~\cite{DBLP:journals/tmi/HooverKG00}, and CHASE\_DB1~\cite{DBLP:journals/tbe/FrazRHUROB12}. DRIVE features 40 colored fundus images in total, and is split 20:20 for training and testing. 1stHO (Human Observer) labels are available for both training and testing, with an additional 20 testing labels from a 2ndHO. All images has a plane resolution of 584 $\times$ 565. CHASE\_DB1 features 28 colored fundus images from 14 patients (left and right eye). Each image has a plane resolution of 960 $\times$ 999, and is annotated by two Human Observers. STARE features 20 colored fundus images, each image has a plane resolution of 960 $\times$ 999, and is also annotated by two Human Observers.

Vessel segmentation models in our experiments were trained on the DRIVE training set exclusively. As there are only 20 images available, models were trained using a 5-fold cross-validation split to assess and ensure its generalizability. In each fold, the checkpoint with the best validation dice score is saved. All images were pre-preprocessed, rescaled to a common resolution of 1024 $\times$ 1024, and cached in RAM for fast access. Data augmentation is performed on-the-fly, and includes; random 360$^{\circ}$ rotation, random left-right flipping, random brightness, and random contrast enhancements.

The architecture selected for the segmentation models is DeepLabV3+~\cite{DBLP:conf/eccv/ChenZPSA18}, from \url{keras.io} examples. Other popular segmentation models, such as XceptionUNet\footnote{\url{https://keras.io/examples/vision/oxford_pets_image_segmentation/}} or ResidualUNet\footnote{\url{https://keras.io/examples/generative/ddim/\#network-architecture}}, can be easily used in-place. The models were trained for 2000 epochs using Adam optimizer, with a batch size of 2 and an initial learning rate of $10^{-4}$. The loss function is a combination of Binary Crossentropy and Soft Dice Loss with equal weighting, as defined in Equation~\ref{eqn:loss}. All experiments were conducted on a computer with Intel Xeon E5-1630 CPU and NVIDIA A5000 GPU, using Tensorflow 2.0 + Keras. Training time takes approximate 1hr35min to reach convergence. 

\begin{equation}
    \mathcal{L}(y,\hat{y}) = 1 - \frac{2 \cdot \sum_{i=0}^{N}y_i \cdot \hat{y_i}}{\sum_{i=0}^{N}(y_i+\hat{y_i}) + \epsilon} 
    -\frac{1}{N}\sum_{i=0}^{N}\big(y_i\cdot\log{\hat{y_i}} + (1-y_i)\cdot\log(1-\hat{y_i})\big)
\label{eqn:loss}
\end{equation}

Table~\ref{tab:results} shows the performance of our model compared to related methods in recent literature, as well as other vessel segmentation datasets. It can be seen that models tested on the DRIVE test set achieved an average dice score of around 0.75, with sensitivity, specificity and AUC of around 0.78, 0.98 and 0.98 respectively. With a CI value in the order of $10^{-3}$, this shows that the models are fairly robust in their predictions. The achieved scores are also higher compared to CHASE\_DB1 and STARE, which is expected as they are in domain. For CHASE\_DB1 and STARE, the models were able to maintain a segmentation performance around 0.7 dice. The CI remains low in the order of $10^{-3}$ for STARE, but is one order of magnitude higher at $10^{-2}$ for CHASE\_DB1. Comparing to related works in literature, our model is also able to perform in-line with the state-of-the-art. 

% Please add the following required packages to your document preamble:
% \usepackage{booktabs}
\begin{table}[ht]
\centering
\begin{tabular}{@{}lcccc@{}}
\toprule
                        & Dice / F1  & Sensitivity   & Specificity   & AUC~~        \\ 
                        \midrule
~~U-Net (2015)~\cite{DBLP:conf/miccai/RonnebergerFB15}
                        & ---        & 0.7537       & 0.9820       & 0.9755~~    \\
~~R2U-net (2018)~\cite{DBLP:journals/corr/abs-1802-06955}
                        & 0.8171     & 0.7792       & 0.9813       & 0.9784~~    \\
~~DUNet (2019)~\cite{DBLP:journals/kbs/JinMPCWS19}          
                        & 0.8203     & 0.7894       & 0.9870       & 0.9856~~    \\
~~NUA-Net (2019)~\cite{DBLP:journals/pr/Zhao0WY21}
                        & ---        & 0.8060       & 0.9855       & 0.9878~~    \\
~~RV-GAN (2021)~\cite{DBLP:conf/miccai/KamranHTZSB21}
                        & 0.8690     & 0.7927       & 0.9969       & 0.9887~~    \\
~~DF-Net (2022)~\cite{DBLP:journals/inffus/YinCW22}
                        & ---        & 0.7689       & 0.9848       & 0.9866~~    \\
                        \midrule
~~DRIVE (1stHO)         & 0.757 (0.001) & 0.795 (0.005) & 0.974 (0.000) & 0.963 (0.000)~~    \\
~~DRIVE (2ndHO)         & 0.766 (0.001) & 0.814 (0.007) & 0.974 (0.001) & 0.968 (0.001)~~    \\
~~CHASE\_DB1 (1stHO)    & 0.699 (0.024) & 0.774 (0.051) & 0.972 (0.004) & 0.955 (0.010)~~    \\
~~CHASE\_DB1 (2ndHO)    & 0.687 (0.021) & 0.756 (0.080) & 0.971 (0.007) & 0.952 (0.013)~~    \\
~~STARE (ah)            & 0.726 (0.004) & 0.724 (0.021) & 0.977 (0.002) & 0.940 (0.004)~~    \\
~~STARE (vk)            & 0.701 (0.007) & 0.662 (0.025) & 0.982 (0.002) & 0.919 (0.008)~~    \\ 
\bottomrule
\end{tabular}
\caption{Quantitative Results of Vessel Segmentation Models. $(\cdot)$ denotes Confidence Interval at 95\%. Top half table: performance of models in related works trained and tested on DRIVE dataset. Bottom half table: performance of models in our experiments that have only been trained on DRIVE dataset.}
\label{tab:results}
\end{table}

\begin{figure}[ht]
    \centering
    \includegraphics[height=1.6cm]{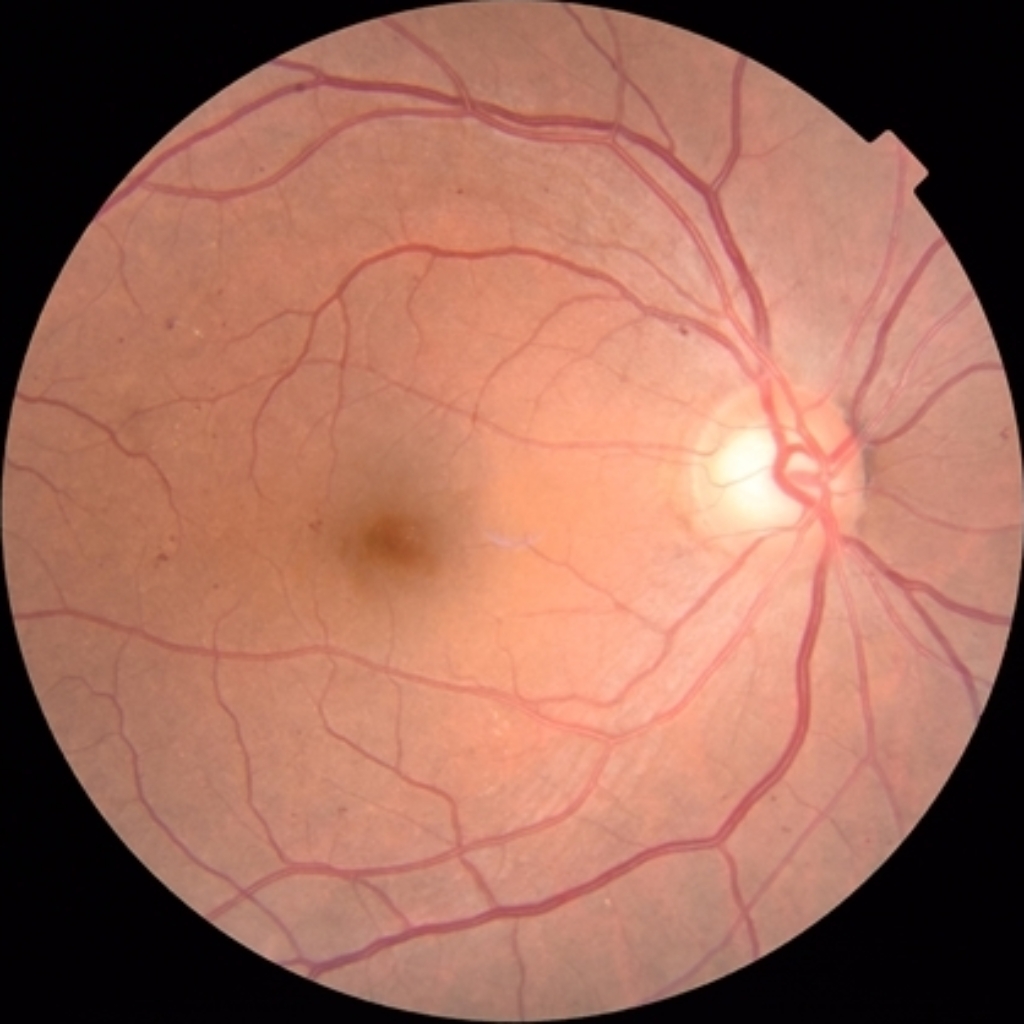} \hspace{1mm}
    \includegraphics[height=1.6cm]{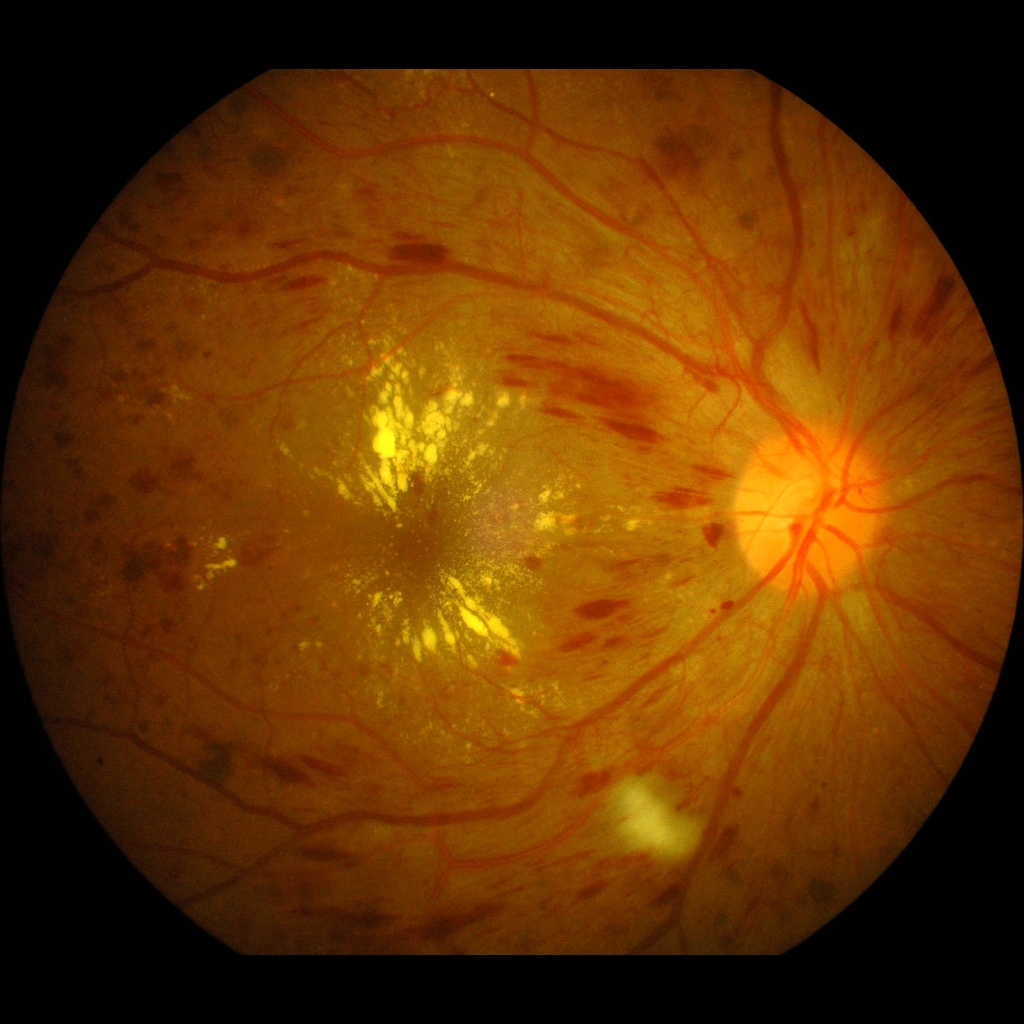} \hspace{1mm}
    \includegraphics[height=1.6cm]{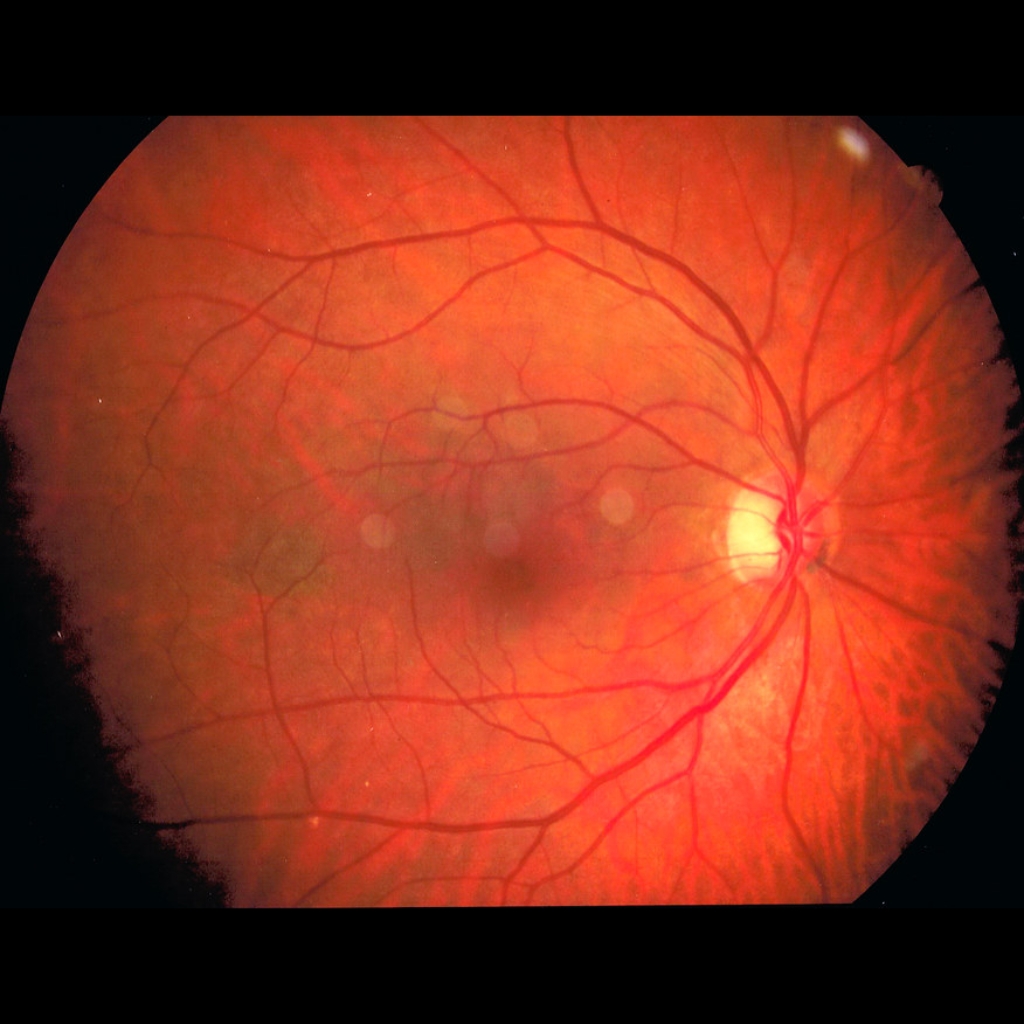} \hspace{1mm}
    \includegraphics[height=1.6cm]{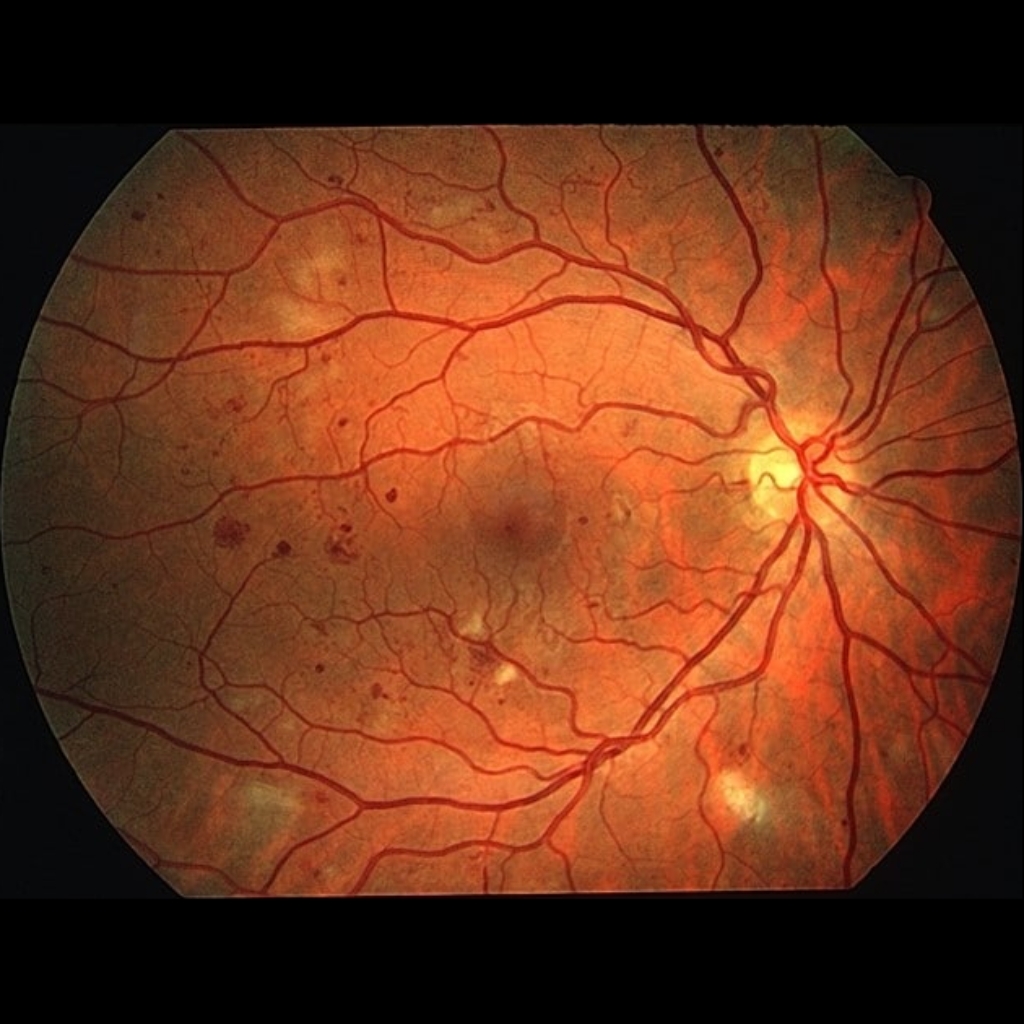} \hspace{1mm}
    \includegraphics[height=1.6cm]{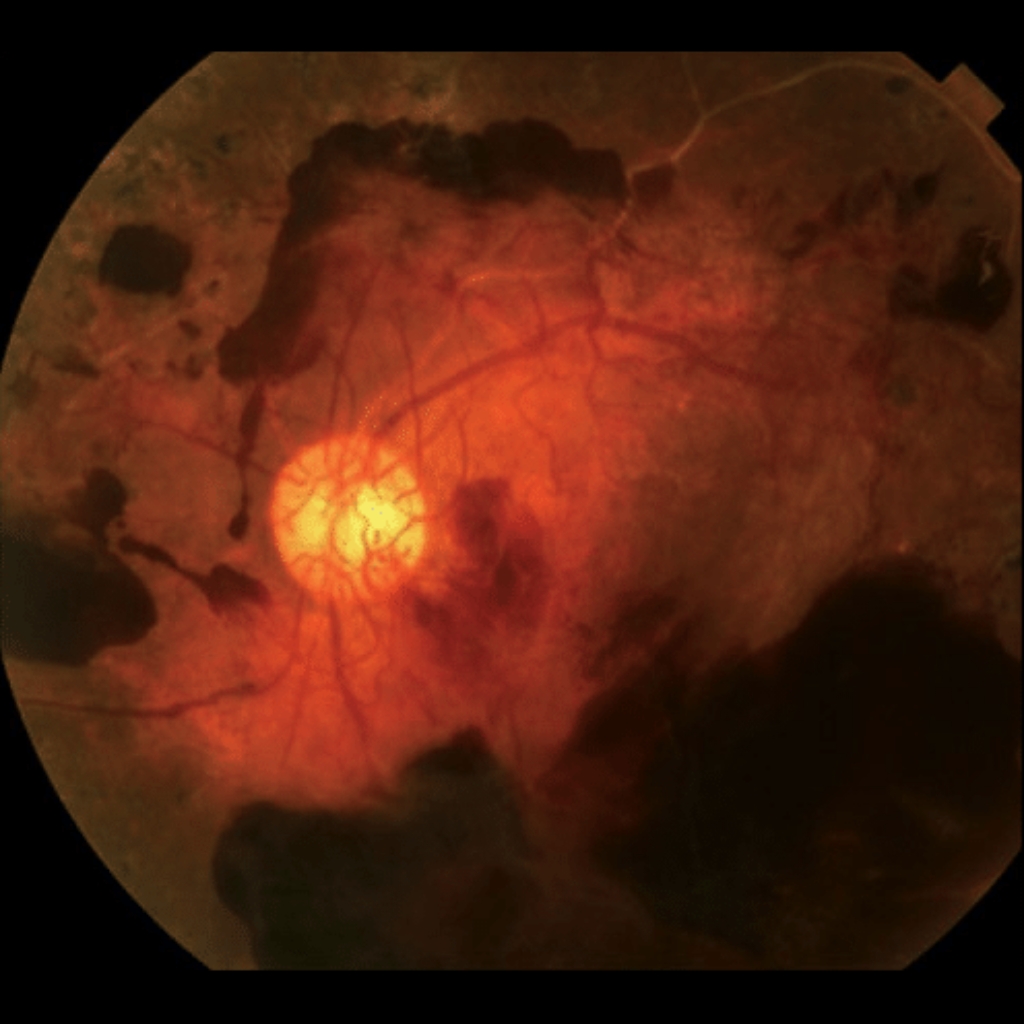} \hspace{1mm}
    \includegraphics[height=1.6cm]{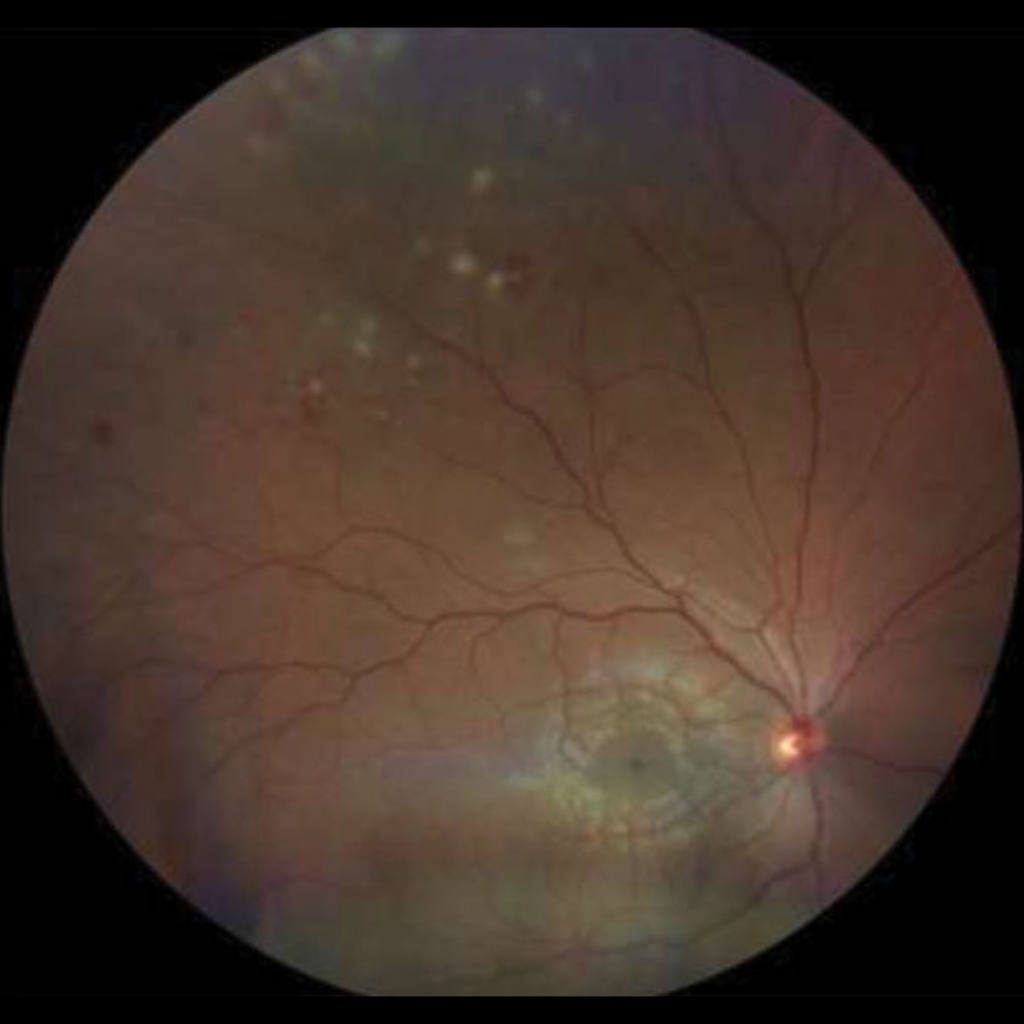}\\
    \includegraphics[height=1.6cm]{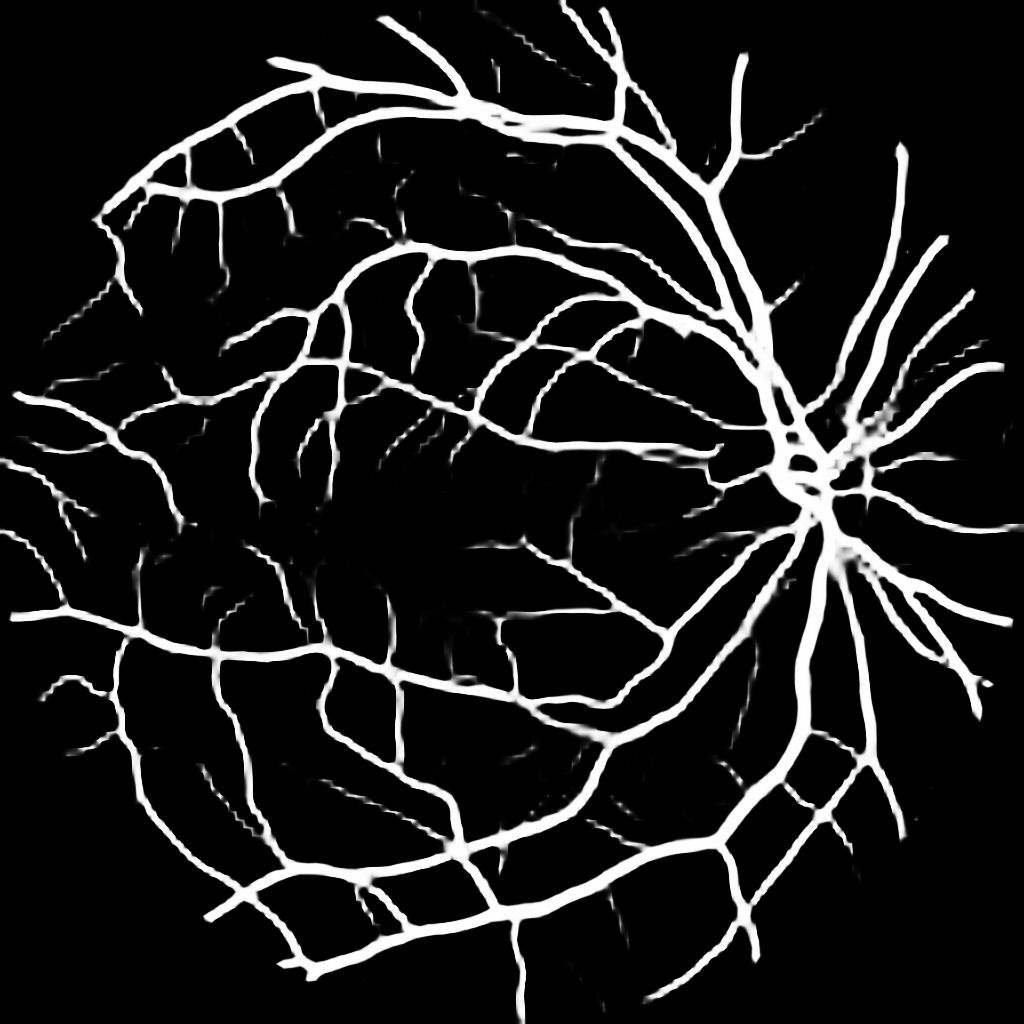} \hspace{1mm}
    \includegraphics[height=1.6cm]{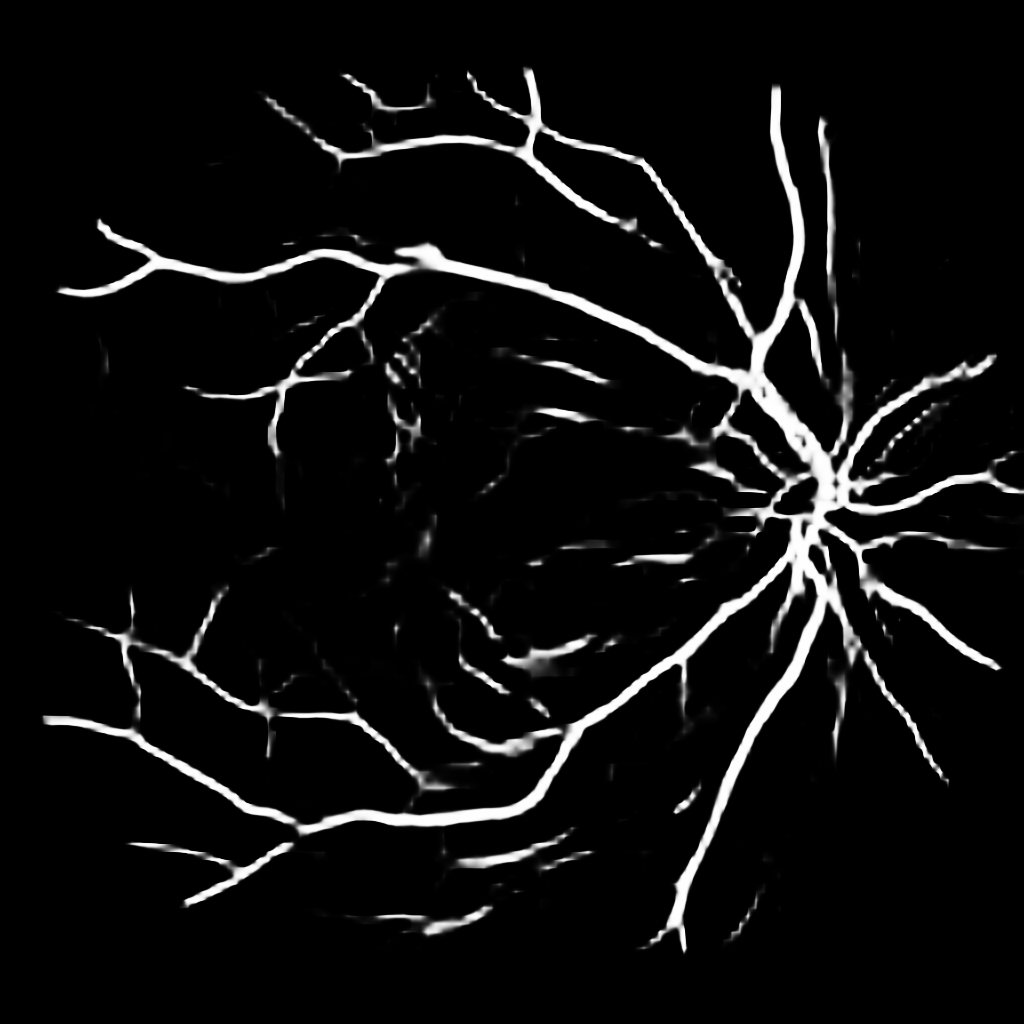} \hspace{1mm}
    \includegraphics[height=1.6cm]{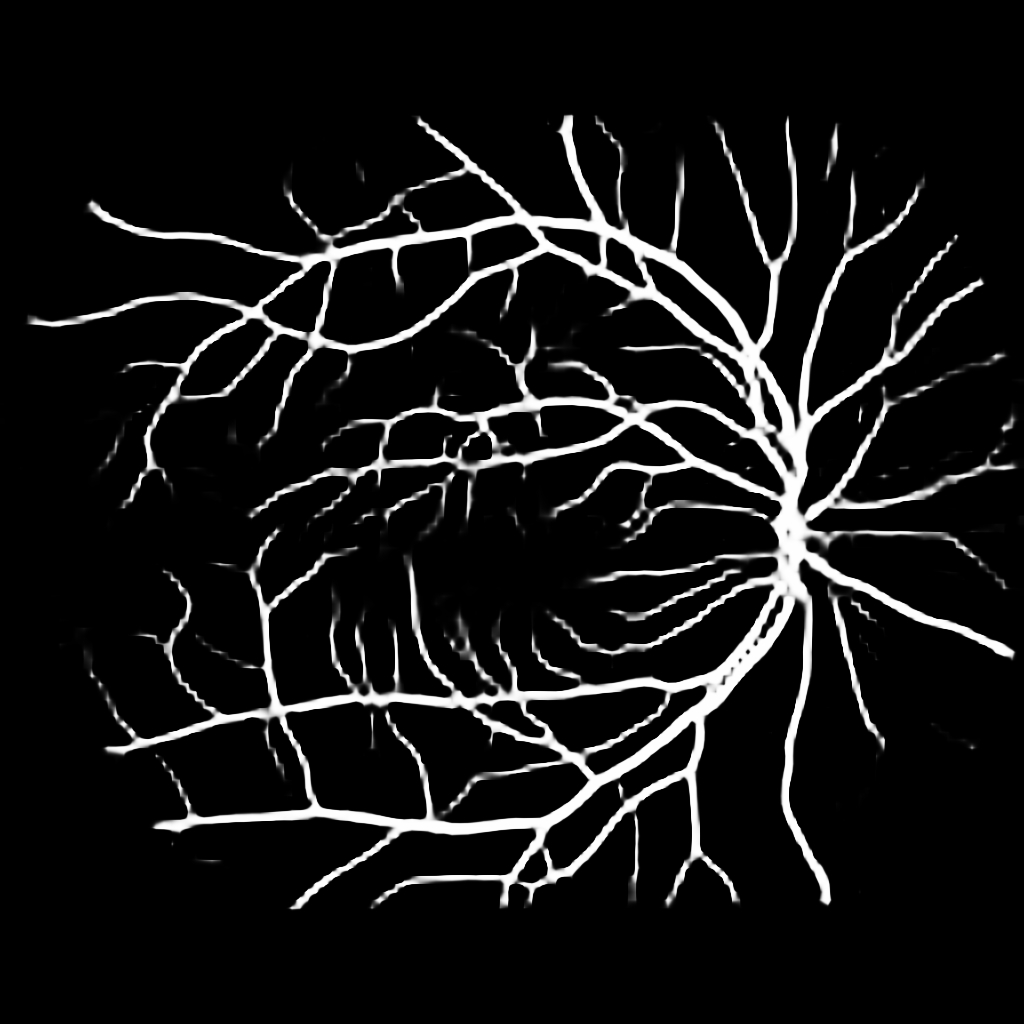} \hspace{1mm}
    \includegraphics[height=1.6cm]{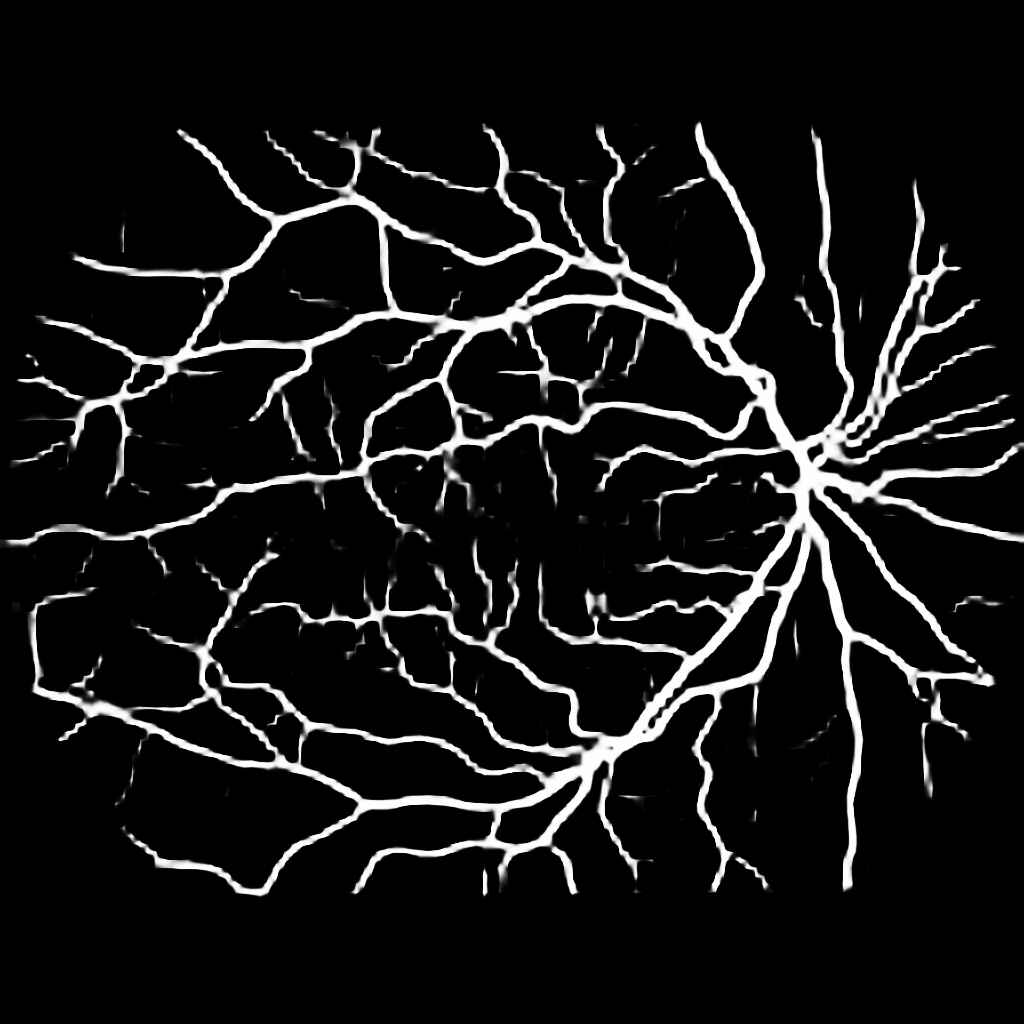} \hspace{1mm}
    \includegraphics[height=1.6cm]{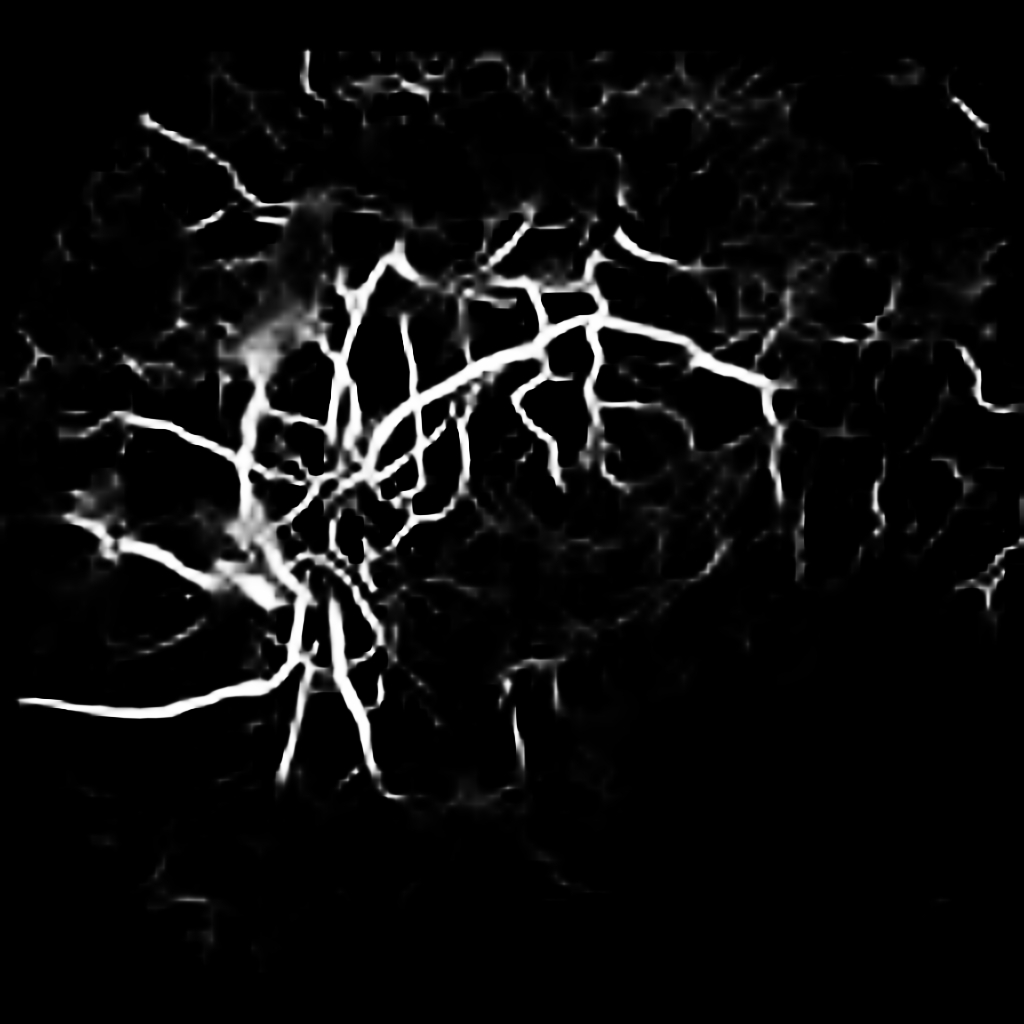} \hspace{1mm}
    \includegraphics[height=1.6cm]{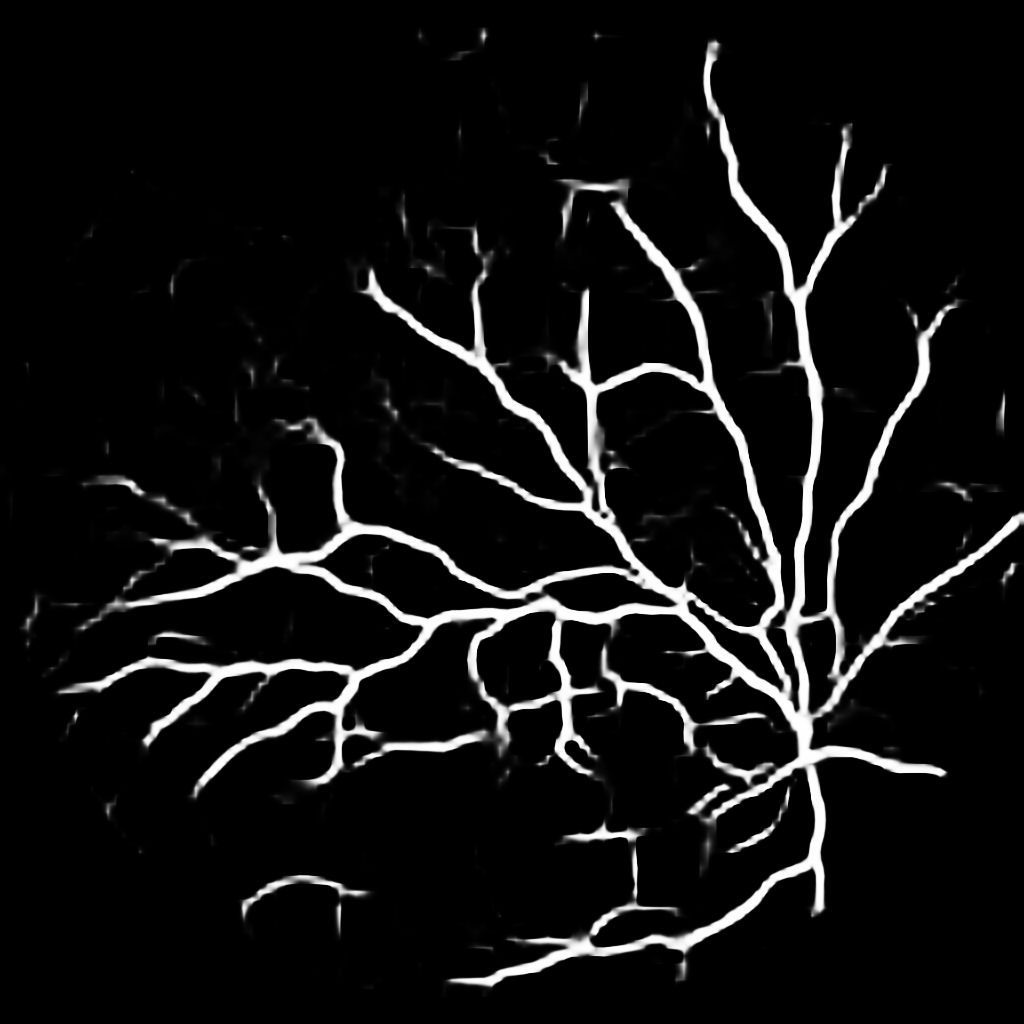}
    \caption{Test images obtained from Google Images} 
    \{L to R: (1) Healthy, (2) Unhealthy, (3) Speckle Artefacts, (4) NPDR, (5) PDR, (6) Hemorrhages\}
    \label{fig:test_images}
\end{figure}

For completeness, and to test the robustness of our method, the models were also tested with several images sourced from Google\texttrademark. We purposefully sourced extremely poor quality images (i.e. images with low contrast, speckle artefacts, etc), as well as images of pathological cases (i.e. Exdudates, Hemorrhages, etc). As there are no ground truth labels, only a qualitative assessment can be made of the results shown in Figure~\ref{fig:test_images}. In all cases, the models have accurately segmented the vessel structure whilst ignoring imaging artifacts and pathological lesions. This is very evident in the case of (5) and (6) where no vessels were segmented in areas of severe hemorrhaging. 

\section{Discussion and Conclusion}

In this paper, we have shown it is possible to achieve near state-of-the-art performance by crafting a well-thought pre-processing pipeline for retina fundus images instead of resorting to complex network architectures. The pipeline locates the spherical retina in irregular fundus photos, and assimilates it onto a regular square canvas. With the spherical retina perfectly centered, it is possible to data augment with random 360$^{\circ}$ rotation, as well as random brightness and contrast enhancements. Models trained through our method are not only just robust across datasets, but also robust against pathological cases. This shows that a less complex and simple approach can generalise better and be more practical. Despite not surpassing state-of-the-art methods, performance may still be gained though tweaking of hyper-parameters. This will be reserved for future work.

\section*{Potential Negative Societal Impact Statement}

The product of retina vessel segmentation is not particularly useful on its own, but plays a crucial role in multiple downstream tasks such as detecting Diabetic Retinopathy, Age-related Macular Degeneration, etc. Segmenting retina vessels is an extremely long and tedious task, and is a very taxing activity if conducted for a prolonged amount of time. With the growing number of eye related illnesses increasing year-on-year, this places a burden on ophthalmologists, especially in places where well-trained ophthalmologists are scarce. Ophthalmologists would benefit from any help they can get, thus we believe the benefit of an AI solution outweighs any Potential Negative Societal Impact. 

As it is impossible to guarantee a perfect performing model, we encourage that a ``vessel segmentation tool'' can give ophthalmologists a good starting point and not to take the prediction by face value. The consequence of a misdiagnosis would be drastic to the patient, as most eye-related illnesses are unidirectional, where early detection and treatment can prevent further degradation. 

\printbibliography

\end{document}